# Generating single longitudinal mode entangled photons in telecom band via a submillimeter monolithic cavity

Yin-Hai Li,[3,1,2,4] Zhi-Yuan Zhou,[1,2,3,4†] Shi-Long Liu,[1,2] Yan Li,[1,2,4] Shi-Kai Liu,[1,2] Chen Yang,[1,2] Zhao-Huai Xu,[1,2] Shuang Wang,[1,2] Zhi-Han Zhu,[3] Wei Gao,[3] Guang-Can Guo,[1,2] Bao-Sen Shi[1,2,3†]

[1] *Key Laboratory of Quantum Information, University of Science and Technology of China, Hefei, Anhui 230026, China*
[2] *Department of Optics and Optical Engineering, University of Science and Technology of China, Hefei, Anhui 230026, China*
[3] *Wang Da-Heng Collaborative Innovation Center for Science of Quantum Manipulation & Control, Heilongjiang Province & Harbin University of Science and Technology, Harbin 150080, China*
[4] *Qingdao Kunteng Applied Quantum Technology Co. Ltd., Qingdao, Shan Dong, 266109, China*

*† Corresponding authors: zyzhouphy@ustc.edu.cn; drshi@ustc.edu.cn*

A high-quality, compact, and narrow-bandwidth entangled photon source (EPS) is indispensable for realization of many quantum communication protocols. Usually, a free space cavity containing a nonlinear crystal is used to generate a narrow bandwidth EPS through spontaneous parametric down-conversion (SPDC). One major drawback is that this occupies a large space and requires complex optical and electrical control systems. Here we present a simple and compact method to generate a single-longitudinal-mode time-energy EPS via type-II SPDC in a submillimeter Fabry-Pérot cavity. We characterize the quality of the EPS by measuring the coincidence-to-accidental coincidence ratio, the two-photon time cross-correlation function and the two-photon interference fringes. All measured results clearly demonstrate that the developed source is of high quality when compared with EPSs generated using other configurations. We believe this source is very promising for applications in the quantum communication field.

An entangled photon source (EPS) is indispensable for many quantum information applications, including quantum communication [1-3], computation [4-6] and metrology [7-9]. Quantum memory is an important component for construction of quantum repeaters [10,11] to realize long-distance quantum communications, where effective interaction between a photon acting as the information carrier and matter (e.g., an atom) acting as the memory is required. However, most quantum memories operate at visible wavelengths [12] and are thus incompatible with the low-loss transmission window of optical fibers (around 1550 nm). To bridge the gap between these wavelength ranges, a quantum interface must be used [13]. Furthermore, for effective coupling to the memory medium, the photon must have a bandwidth that is comparable with the natural bandwidth of an atom, which is generally less than 100 MHz [14,15]. Therefore, generation of a high-quality, narrow-bandwidth EPS has long been pursued as a target for quantum communication.

Spontaneous parametric down-conversion (SPDC) in a nonlinear crystal is a commonly used method for EPS generation [16-19], but the emission spectrum in a single-pass configuration is usually too broad (~THz) to match the linewidth of a typical quantum memory (~100 MHz) [14]. However, when a nonlinear material is inserted into a cavity, the bandwidth of a photon emitted from that cavity is determined by the linewidth of the cavity and can be reduced significantly. In addition, the spectral brightness can also be enhanced when compared with that of the single-pass configuration. Since the pioneering work presented in Ref. [20], considerable progress has been made in this area [20-23]. Initially, most cavity-enhanced SPDC sources were used to generate degenerate single-longitudinal-mode photon pairs [24]. More recently, two-color narrow-bandwidth photon pairs with one photon in the visible band (corresponding to the wavelength of a specific quantum memory) and the other photon in the telecommunications band [15,21] were generated. Two types of cavity configurations are used for narrow-bandwidth photon pair generation: in the first configuration, only two down-converted photons are resonant with the cavity [25]; in the second configuration, both the pump and the down-converted photons are resonant with the cavity [20,26]. In almost all reported sources for narrow-band photons, a free-space cavity with a nonlinear crystal located inside it is used, which occupies a huge area on the optical table and requires complex auxiliary optical and electrical control systems; this makes the source difficult to operate and it can be challenging to build a compact and stable system. To simplify the experimental setup and minimize the size of the photon source, Chuu *et al.* reported generation of a 1064 nm narrow-bandwidth photon pair using a 1-cm-long monolithic cavity [25].

In this work, we report on generation of a single-longitudinal-mode narrow-bandwidth time-energy EPS in the telecommunications band for the first time using a submillimeter Fabry-Pérot (FP) cavity fabricated by direct coating on a thin type-II periodically poled potassium titanyl phosphate (PPKTP) crystal. We obtain a maximum coincidence to accidental-coincidence counts ratio (CAR) of 1800, and the measured cross-time-correlation between the photon pair yields bandwidths for the two orthogonally polarized photons of 546 MHz and 735 MHz. Two-photon Franson interference gives a raw (net) visibility of more than 87.12% (88.58%). The estimated spectral brightness is 2.636 (s·mW·MHz)$^{-1}$. A single-photon Michelson interference experiment provides further clear verification of the single longitudinal mode characteristics.

The experimental setup is illustrated in Fig. 1(a). A PPKTP crystal with two end faces (along the pump propagation direction, the front face is coated with an anti-reflection (AR) coating at 775 nm and a highly reflective (HR) coating at 1550 nm, while the back face is coated with an AR coating at 775 nm and a partially-reflective (PR) coating at 1550 nm) forms an FP cavity that is resonant at 1550 nm. The crystal has dimensions of 1 mm×2 mm×0.85 mm and is periodically poled with periodicity of 46.2 μm to obtain quasi-phase matching for SPDC with a pump beam at 775 nm and signal and idler photons at 1550 nm. The crystal temperature is controlled using a homemade temperature controller with stability of ±1 mK. The pump beam propagates along the *x* axis. The crystal is designed for type-II phase matching of $n_y(\omega_p) \rightarrow n_y(\omega_s) + n_z(\omega_i)$, where $n_{y,z}$ represents the refractive indices along the *y* and *z* axes of the crystal and $\omega_{p,s,i}$ represents the frequencies of the pump, signal, and idler fields, respectively. A 775 nm continuous wave laser (TA Pro, Toptica, Munich, Germany) is used as the pump. The beam shape is modulated using a group of lenses so that the photons generated by SPDC are resonant with the cavity. A dichroic mirror is used here to verify the cavity status via a photoelectric detector and an oscilloscope (OSC) (see supplementary, section 1). After the PPKTP crystal, an infrared lens is used to collect the photons with high efficiency. The signal and idle photons are then divided using a polarizing beam splitter (PBS) and collected, while the pump beam is removed by filters (FELH 1400/1000, Thorlabs, USA). The signal and idle photons are sent to all-fiber-based unbalanced Michelson interferometers (UMIs, each consisting of a fiber beam splitter and two Faraday rotation mirrors) to generate the time-energy entanglement. Two superconductor nanowire single-photon detectors (SNSPDs, Scontel, Russia) and a coincidence device (Timeharp 260, PicoQuant, Germany) are used to carry out the single-photon and two-photon correlation measurements.

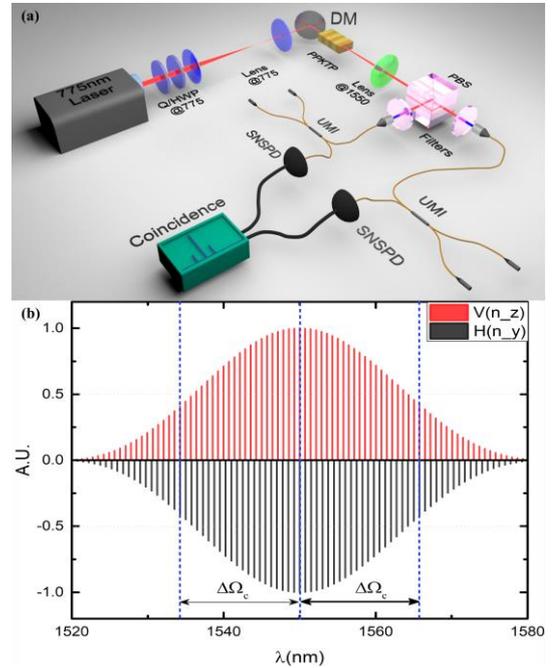

FIG. 1. (a) Experimental setup. Q/HWP: quarter/half wave plates; DM: dichroic mirror; highly reflective (HR) @775 nm and highly transmissive (HT) @1550 nm; PPKTP: periodically poled potassium titanyl phosphate; PBS: polarizing beam splitter; Filters: long pass filters; UMI: unbalanced Michelson interferometer, consisting of one fiber beam splitter and two Faraday rotation mirrors; SNSPD: superconductor nanowire single-photon detector.

SPDC should satisfy the conservations of both energy ($\omega_p = \omega_s + \omega_i$) and momentum ($k_p = k_s + k_i$) [27]. Using the quasi-phase matching condition, we calculate the spectrum and the bandwidth of the SPDC shown in FIG. 1(b). According to the formula $\Delta\omega = c/(nl)$ [20], the mode spacing $\Delta\omega_i$ of the idle frequency is slightly less than the mode spacing $\Delta\omega_s$ of the signal (where $n_z > n_y$), where $\Delta\omega$ is the free spectral range (FSR) of the cavity, $n$ is the refractive index, and $l$ is the optical length. At the blue dashed lines shown in Fig. 1(b), both the signal and the idle modes are resonant. We call the frequency spacing of these doubly resonant modes the cluster spacing (denoted by $\Delta\Omega_c$) [26], which can be expressed as:

$$\Delta\Omega_c = \frac{\Delta\omega_s \cdot \Delta\omega_i}{\Delta\omega_s - \Delta\omega_i} \quad (1)$$

With a frequency shift of $\Delta\Omega_c$ from the central blue dashed line, i.e., at the left and right blue dashed lines (Fig. 1(b)), both the signal and idle modes will become resonant again. The number of cavity modes that a cluster spacing can contain for the signal and idle photons can be expressed as:

$$N_s = n_y/(n_z - n_y), N_i = n_z/(n_z - n_y) \quad (2)$$

Obviously, $N_i - N_s = 1$. Generally, $N_s$ and $N_i$ are not integers, i.e., $Mod(N_{s,i}, 1) \neq 0$. Therefore, at the positions of the left or right blue dashed lines, there are frequency differences between the orthogonal cavity modes:

$$\Delta\nu = \begin{cases} Mod(N_{s,i},1) \times (\Delta\omega_s - \Delta\omega_i), Mod(N_{s,i},1) < 1/2 \\ (1 - Mod(N_{s,i},1)) \times (\Delta\omega_s - \Delta\omega_i), Mod(N_{s,i},1) > 1/2 \end{cases} \quad (3)$$

Given that the bandwidths of the signal $\Delta\nu_s$ and the idler $\Delta\nu_i$ are sufficiently narrow, i.e., $\Delta\nu_s + \Delta\nu_i < \Delta\nu$, we will only obtain orthogonal modes that are doubly resonant at the central blue dashed line position, which means that a single-longitudinal-mode photon pair is generated. This conclusion is also applicable to nondegenerate cases.

We first characterize the FP cavity with an infrared laser source (CTL 1550 nm, Toptica, Germany). The detailed experimental setup is presented in the Supplementary Materials. The results obtained are shown in FIG. 2.

According to equations (1), (2) and (3), we have $\Delta\Omega_c = 1997.75$ GHz, $N_s = 21.34$, $N_i = 22.34$ and $\Delta\nu = 1.43$ GHz $> \Delta\nu_s + \Delta\nu_i = 1.281$ GHz. The results fit the single longitudinal mode theory well.

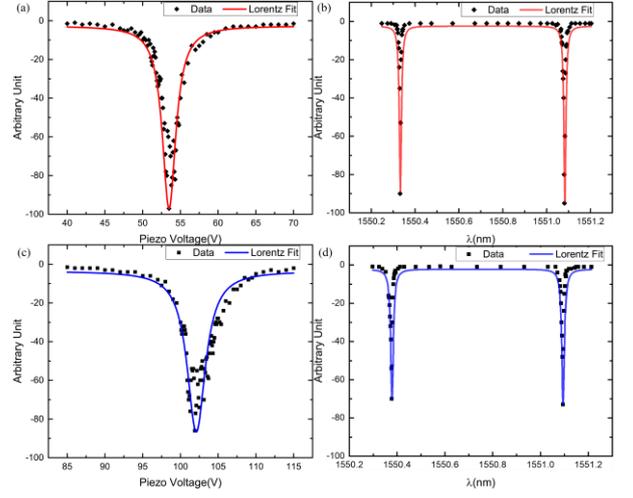

FIG. 2. Characterization of the FP cavity. (a) The bandwidth of the horizontal-polarization cavity mode is 546 MHz. (b) The FSR of the horizontal-polarization cavity mode is 93.61 GHz. (c) The bandwidth of the vertical-polarization cavity mode is 735 MHz. (d) The FSR of the vertical-polarization cavity mode is 89.42 GHz.

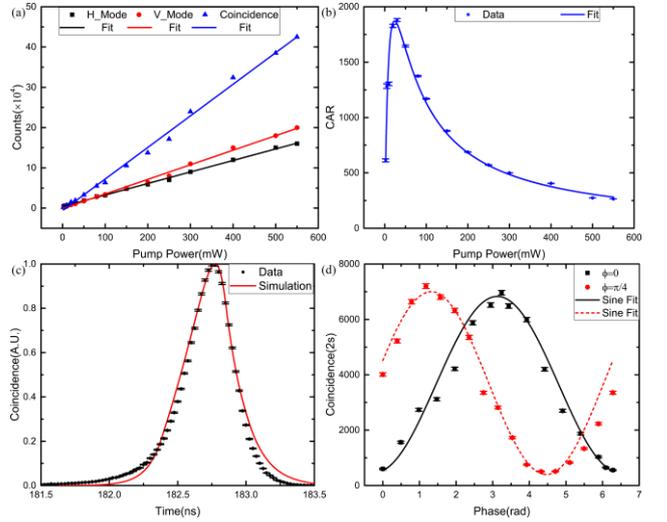

FIG. 3. Illustrations of correlated photon pairs and interference curves of the time-energy entanglement. (a) Single count (black squares for signal photons, red circles for idler photons) and coincidence count (blue triangles) vs. pump power and linear fits. (b) Coincidence to accidental-coincidence ratio (CAR) vs. pump power. (c) Glauber correlation functions of the signal and idler photons. Coincidence counts (black squares) are measured as a function of the time delay between the signal and idler photons at a pump power of 300 mW. The blue solid line is the result of a theoretical simulation of equation (5). (d) Interference curves: coincidence counts and fittings. The coincidence counts are measured as a function of the phase of the idler photons. Red circles represent coincidence counts when the UMI phase of the idle photons is at $\pi/4$, while black squares represent the coincidence counts when the UMI phase of the idle photons is at phase 0.

We then perform the correlation measurements. First, we measure the single counts and the coincidences at various pump powers without the UMIs (with results as shown in FIG. 3(a)). The heralded efficiency is approximately 25%.

The collection efficiency for single photons is approximately 55%. We then calculate the CAR (shown in FIG. 3 (b)) using the equation $CAR = (R_c + R_{ac})/R_{ac}$ [28], which has values of approximately 500 and 1800 at pump powers of 300 mW and 50 mW, respectively. Here, $R_c$ is the coincidence count, and $R_{ac}$ is the accidental coincidence count, which is mainly a result of the noise from the detector. We estimate the photon pair generation rate to be 2.636 (s·mW·MHz)$^{-1}$ using the formula $R_{estimate} = R_{detected}/(d\alpha_1\alpha_2 t_1 t_2 \eta^2)$ [20], where $R_{estimate}$ is the estimated pair production rate, $R_{detected}$ is the detected pair rate, $\alpha_{1,2}$ and $t_{1,2}$ are the fiber collection efficiency and transmittance of the filters, respectively, for the signal and idler photons, $\eta$ is the detection efficiency of the SNSPDs and $d$ is the detection duty cycle of each trigger period. For the SNSPDs, $d = 1$.

The Glauber correlation function measurements are performed with 25 ps coincidence resolution and 10 s coincidence times at a pump power of 300 mW. The Glauber correlation function for type-II double resonant SPDC is given as [20,29]:

$$G_{s,i}^{(2)}(\tau) = \langle \Psi | E_i^{(-)}(t) E_s^{(-)}(t+\tau) E_s^{(+)}(t+\tau) E_i^{(+)}(t) | \Psi \rangle$$
$$\propto \left| \sum_{m_s,m_i=-\infty}^{\infty} \frac{\sqrt{\gamma_s \gamma_i \omega_s \omega_i}}{\Gamma_s + \Gamma_i} \begin{cases} e^{-2\pi\Gamma_s(\tau-\tau_0/2)} \operatorname{sin} c(i\pi\tau_0 \Gamma_s) \ \tau \geq \tau_0/2 \\ e^{2\pi\Gamma_i(\tau-\tau_0/2)} \operatorname{sin} c(i\pi\tau_0 \Gamma_i) \ \tau < \tau_0/2 \end{cases} \right|^2 \quad (4)$$

where $E_s^{(\pm)}$ and $E_i^{(\pm)}$ are the operators for the signal and idler photons, $\gamma_{s,i}$ are the cavity damping rates for the signal and idler modes, and $\omega_{s,i}$ are the central frequencies for the signal and idler photons, respectively; $\tau_0$ is the propagation delay between the signal and idler photons inside the crystal, and $\Gamma_{s,i} = \gamma_{s,i} + im_{s,i}\Delta\omega_{s,i}$ with mode indices $m_{s,i}$ and a free spectral range $\Delta\omega_{s,i}$. The SNSPD's response can be modeled as $\phi(t) = \alpha e^{\gamma t/2}\Theta(-t)$ [30], where $\Theta(-t)$ is the Heaviside step function and $\gamma$ is the damping rate. Therefore, the measured coincidence count curve is the convolution of $G_{s,i}^{(2)}(\tau)$ and $\phi(t)$:

$$\Gamma(t) = G_{s,i}^{(2)}(\tau) * \phi(t) \quad (5)$$

The measured Glauber correlation function is shown in FIG. 3(c) as black squares. The blue solid line represents the numerical simulation of equation (4) and this line fits the measurement results quite well. The full width at half maximum (FWHM) of the Glauber correlation function is approximately 0.412 ns. Equation (4) shows that the correlation function has oscillatory damping with a multipeak comb-like shape, where the damping rate of the peak values is equal to the cavity damping rate. The envelope of these peaks decays as a function of $e^{-2\pi\gamma_{s,i}|\tau|}$. In our cavity, $\gamma_s$ and $\gamma_i$ are 546 MHz and 735 MHz, respectively. Without consideration of the responses of the detection and coincidence systems, the FWHM of the Glauber correlation time is defined as $T_{FWHM} = 1.39/2\pi\gamma$, where $\gamma$ is the geometric mean of $\gamma_s$ and $\gamma_i$. Therefore, we should obtain an FWHM of 0.349 ns. In the experiments, the detection and coincidence systems resulted in the broadening of the time cross-correlation curve to 0.412 ns.

We then use two UMIs to generate the time-energy entanglement. We measure the coincidences by tuning the UMI phase of the signal photons while the phase of the idler photons remains fixed. The results are shown in FIG. 3(d). The visibilities are (87.12±0.711)% ((88.58±0.656)%) and (89.19±0.634)% ((90.66±0.604)%) without (with) subtraction of the accidental coincidence counts when the idle UMI phase is at 0 and at π/4, respectively. The details of the UMI phase tuning are presented in the Supplementary Materials.

To verify the single longitudinal mode feature of the photons, we perform a single-photon Michelson interference experiment. The experimental setup and the results are illustrated in FIG. 4.

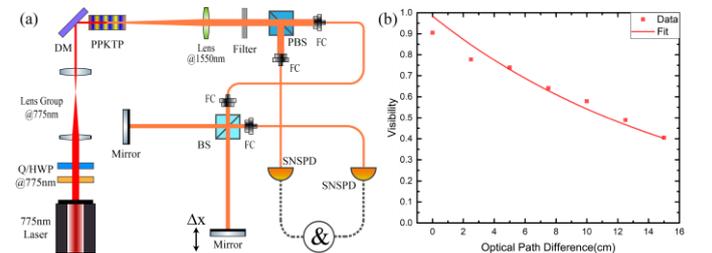

FIG. 4. (a) Experimental setup. Q/HWP: quarter/half wave plates; DM: dichroic mirror; PPKTP: periodically poled potassium titanyl

phosphate; Filter: long pass filter; PBS: polarizing beam splitter; FC: fiber collimator; BS: beam splitter; SNSPD: superconductor nanowire single-photon detector. (b) Experimental results. Red squares represent the visibility as a function of the optical path difference $\Delta x$. The red solid line is the best fit.

As shown in FIG. 4(a), the horizontally polarized photons (i.e., the signal photons) are directed into a free-space Michelson interferometer and the output is coupled into a SNSPD to carry out coincidence measurements with the idle photons. At different values of $\Delta x$, we calculate the visibility based on the maximum and minimum values of the coincidence, which are presented in Fig. 4(b). A simulation is also performed.

Consider a Lorentz shape for the cavity spectrum that can be expressed as:

$$\phi(\nu) = \frac{\Delta \nu_N}{2\pi} \frac{1}{(\nu-\nu_0)^2 + (\Delta \nu_N/2)^2} \quad (6)$$

where $\Delta \nu_N$ is the spectral bandwidth and $\nu_0$ is the central frequency. The normalization condition is $\int_{-\infty}^{\infty} \phi(\nu) d\nu = 1$. The intensity of the spectrum is given by $I(\nu) = I_0 \phi(\nu)$. We then obtain the interference intensity of the Michelson interferometer:

$$I_L = 2I_0 + 2I_0 \exp[-\pi \left|\frac{\Delta \nu_N L}{c}\right|] \cos\left(\frac{2\pi \nu_0 L}{c}\right) \quad (7)$$

where $c$ is the speed of light in a vacuum. $L$ is the optical path difference. Assuming a constant value of $R$ as the background for the coincidence, the visibility can then be expressed as:

$$\gamma_L = \exp[-\pi \left|\frac{\Delta \nu_N L}{c}\right|] / (1 + R/2) \quad (8)$$

In the simulations, we evaluate $R$ to be 1/30, corresponding to a coincidence background of approximately 50 in the experiment. From the fitting results, we obtain a $\Delta \nu_N$ of 568.9 MHz, which fits well with the results of the cavity status measurements (546 MHz, with approximately 4.19% deviation).

Finally, we provide an estimate of the overall detection efficiencies for the signal and idler photons. The transmittances of the two filters (FELH 1000 and FELH 1400) are 97% and 99%, respectively. The total fiber transmission loss is 2.5 dB. The detection efficiency of our SNSPD is approximately 60%. The other losses caused by the infrared lens and the PBS are thought to be 1%.

In conclusion, we experimentally realized a single-longitudinal-mode correlation photon pair source using a type-II SPDC process in a short F-P cavity. The cluster spacing and the SPDC gain bandwidth are calculated. We estimated a photon pair generation rate of 2.636 (s·mW·MHz)$^{-1}$. We measured the Glauber correlation function and performed a numerical simulation that fitted the measurement results very well. The single longitudinal mode feature was verified in the single-photon Michelson interference experiment. In addition, we generated a time-energy entanglement. A single-longitudinal-mode photon can couple efficiently with quantum memory systems and will overcome the group velocity dispersion to achieve large transmission distances. Our protocol also shows great potential for use in integrated optics. This work provides a potential roadmap toward generation of a single-longitudinal-mode correlated photon pair source via type-II SPDC in integrated optics, which will hold considerable potential for quantum memory and communication systems applications.

**Acknowledgments**
This work was supported by the National Natural Science Foundation of China (NSFC) (61435011, 61525504, 61605194); the National Key Research and Development Program of China (2016YFA0302600); the Anhui Initiative in Quantum Information Technologies (AHY020200); the China Postdoctoral Science Foundation (2017M622003, 2018M642517); and the Fundamental Research Funds for the Central Universities. We thank David MacDonald, MSc, from Liwen Bianji, Edanz Editing China, for editing the English text of a draft of this manuscript.

# Supplementary materials

**Characterize status of the cavity and calibration of the infrared laser**

Before we perform the time-energy entanglement measurement, we characterize status of the cavity through the SHG process. The details are shown in Fig. S1.

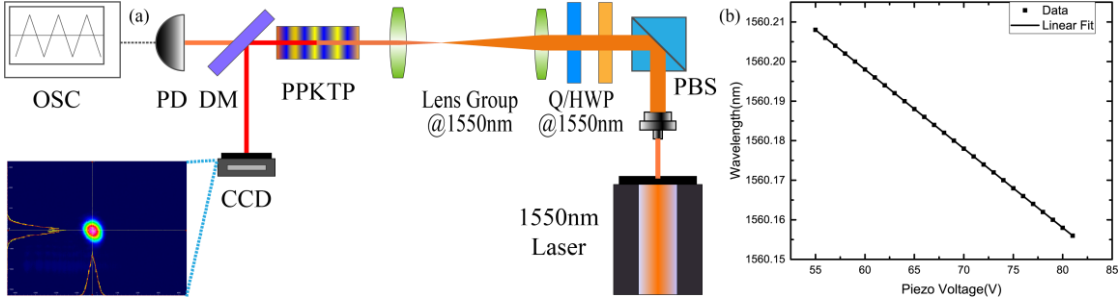

Figure S1. Characterize status of the cavity and calibration of infrared laser. PBS: polarization beam splitter; Q/HWP: quarter/half wave plate; PPKTP: periodic polarized potassium titanium oxide phosphate; DM: dichroic mirror; PD: photoelectric detector; OSC: oscilloscope; CCD: charge-coupled device. On the bottom left is the CCD figure of light generated by SHG process.

We employ an infrared laser (Toptica, CTL 1550nm, German) as the SHG pump. We firstly perform a calibration of the infrared laser, shown in Figure s1 (b). The output wavelength of laser has a good linear relationship with PZT voltage. The quarter and half wave plates adjust polarization of light to horizontal and vertical direction to measure transmission spectrum of the cavity. By fine tuning the PZT voltage i.e. the pump wavelength, we measure the FSR and bandwidth using a PD and OSC. Results are shown in the experimental results section. Then we adjust the polarization of the light be 45 degree with respect to the x axis to satisfy the SHG conditions. The beam is directed into the crystal after a group of lens and a CCD is used to get the SHG beam, as shown in Figure s1 (a), at the left of bottom.

**How UMI phase is tuned**

In the fiber UMI, the thermal coefficient of fiber at 1550 nm is $\frac{dn}{dt} = 0.811 \times 10^{-5} / °C$, the fiber length difference of the UMI is 1.022m, corresponding to $L_d = c\Delta t / 2n$ for 10 ns time delay. The temperature of one tuning period is $\Delta T = \lambda / (2L_d \frac{dn}{dT}) = 0.094 K$. In the experiment, the temperature tuning periods and the phase are calibrated using a stable narrow bandwidth laser source. The phase of the UMIs can keep unchanging for hours because of seriously thermal and acoustic isolation from the environment.